\documentclass[12pt]{article}
\usepackage{amsmath}
\usepackage{mathtools}
\usepackage{amssymb}
\usepackage{amsfonts}
\usepackage{graphicx}

\title{\vspace{-4.0cm} Wavefront sensing based on the inverted Hartmann sensor}
\date{\vspace{-1.0cm} November, 2019}
\begin{document}
\maketitle
\begingroup  
\noindent Job Mendoza-Hernandez$^\ast$ and Dorilian Lopez-Mago
\\
{\footnotesize 
Tecnologico de Monterrey, Escuela de Ingenier\'{i}a y Ciencias, Ave. Eugenio Garza Sada 2501, Monterrey, N.L., M\'exico, 64849.\\
$^\ast$ job.mendoza@alumno.buap.mx}
\endgroup

\section*{Abstract}
The classic Hartmann  test consists of an array of holes to reconstruct the wavefront from the local deviation of each focal spot, and Shack-Hartmann sensor improved that with an array of microlenses. This array of microlenses imposes practical limitations when the wavefront is not into of visible wavelengths, e.g., the fabrication of these. Instead, we propose a wavefront-sensing technique using an array of circular obstructions, i.e.,  the  Hartmann sensor with an inverted Hartmann array. We show that under the same conditions the inverted  Hartmann sensor and the Shack-Hartmann wavefront sensor have an equivalent spot-map to recover the wavefront. The method might  used in non-visible wavelengths and in a wider range of applications.

\section{Introduction}
\label{S:1}

Classically, light is described by a complex function $\tilde{U}$, which carries information about its energy and momentum through its amplitude 
 $|\tilde{U}|$ and phase $\mathrm{arg}(\tilde{U})$, respectively. Measuring the phase distribution at optical frequencies is a challenging task. Since optical detectors measure the time-averaged power ($\propto |\tilde{U}|^{2}$), the phase information is lost in the measurement. There are several techniques to measure the phase distribution. Direct techniques include the Hartmann or Shack-Hartmann (SH) method \cite{Geary}, while indirect techniques are based on the interference pattern between the target beam and a reference beam. 

The measurement and generation of a phase distribution is important in many applications, such as astronomical observations and optical testing\cite{malacara}. For example, in adaptive optics is regularly implemented to compensate wavefront distortion from the atmosphere \cite{Hecht, Becker}. Nowadays, structured light beams containing orbital angular momentum have potential applications in free-space communications \cite{Willner,Willner1}. It has been shown that the phase distribution of such beams can be extracted using interference techniques or a SH wavefront sensor \cite{padgett,gbur}. 
We focus our attention in the Hartamann and SH techniques that measures the shape of a wavefront using an array of holes and microlenses. Each holes and microlens select a section of the wave front at a plane, and it is measure with the position of each spot in other plane (the focal plane in SH). Displacement of each spot, with respect to the optical axis of each microlens or holes, represents the local inclination of the wavevector\cite{Hecht,Shannon}.  A main difference between Hartamann and SH techniques are the possibility of distinguish the angle of spots deviation in the same area \cite{spiricon}.

The fabrication of microlenses for SH sensor most typically is in the visible wavelength range; however, in other range of wavelength is difficult or it is not possible build lenses such as into the wavelength of the extreme ultraviolet (EUV)\cite{polo}, in x-ray \cite{Bakken}, or  Terahertz \cite{Cui}. Hartmann technique is used in Terahertz  \cite{Cui}, and in X-ray range is used a spherical obstruction  \cite{Bakken}. Therefore, there is interest in improving the algorithms to reconstruct the wavefront from the spot-map instead the fabrication of lenses \cite{polo}. Furthermore, the SH technique can be used in the quantum regimen to study optical coherence properties \cite{Stoklasa}. 

The intensity distribution of an individual lens is described by the Airy spot, which is generated by the diffraction of a circular aperture and the lens. The size of this spot is very important in the SH device, since it is possible generated spot-map in order to recover the wavefront; however, there are other factors that can affected of the reconstruction of the wavefront, e. g., the   numerical aperture (NA) of individual lenses, the detector pixel size or the algorithms used for the spot detection \cite{spiricon,polo}, but the spot-map generated by the wavefront is a  fundamental part consideration for measure it.

 We study the feasibility of a wavefront sensing technique that uses an array of circular obstructions, instead of an array of holes or microlenses. The technique uses the Arago spot in order to obtain a 
 spot-map to recover the wavefront \cite{Hecht,wolf,Harvey,bruns}.  Arago spot sometimes is known in the literature as Poisson spot \cite{Lucke}. Circular obstructions can be possible built with lithography methods where the geometric structures or complex structures can be carry out in micrometers or nanometers  \cite{Pimpin,Xia}. Our proposal might using in non-visible wavelength because it is not depending of constructing of lenses, and the Arago spot is generated by the diffraction of field in circular obstructions.

In this work, we demonstrate that using the Arago spot is possible to obtain a discrete spot-map similar to the SH technique, in order to retrieve the wavefront. We make a numerical comparison between the  array of circular obstructions and the SH technique. We study its performance to detect cylindrical, spherical and helical wavefronts.

\section{Airy and Arago spots}

\begin{figure}[ht]
\centerline{\includegraphics[height=5.5cm, width=13cm]{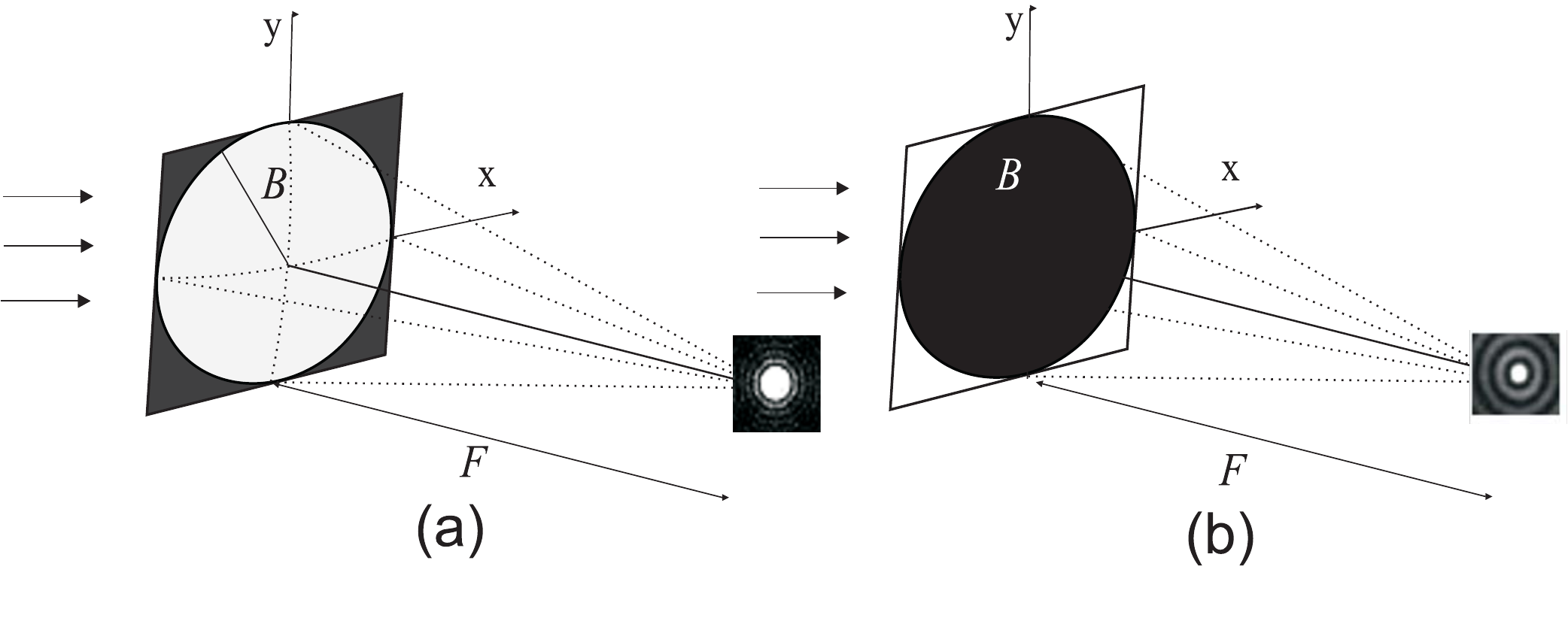}}
\caption{Schematic of the diffraction through (a) a circular thin lens, and  (b) an opaque circular obstruction.}
\label{Fig:1}
\end{figure}

We start by comparing the Fraunhofer pattern of a plane wave diffracted by a circular thin lens of radius $B$ with the diffraction due to a circular obstruction of the same radius. We compare the intensity distribution of both diffraction patterns at the focal point of the lens ($z=F$). Figure \ref{Fig:1} shows the geometry of the problem.

\newpage

The intensity distribution of the Arago spot at the focal plane is described by the function \cite{Harvey,Hecht,wolf}
\begin{equation}\label{sarago}
    I_{Arago}(2BR/F) = I_{0} J_{0}^{2}(2BR/F).
\end{equation}
Similarly, the Airy disk is given by
\begin{equation}\label{sairy}
    I_{Airy}(2BR/F) = I_{1} \frac{J_{1}^{2}(2BR/F) }{(2BR/F)^{2}},
\end{equation}
where $B$ is the radius of the lens, $F$ is the focal distance. $J_{0}$  and $J_{1}$ are the Bessel functions of zero and first order, respectively, and $R$ is the radial coordinate. $I_{0}$ and $I_{1}$ are the maximum amplitudes for each distribution. 
It is important to notice that in the focal plane the intensity in the Airy  spot is due to whole aperture of radius $B$, and the intensity of Arago spot is the diffraction by the circular obstruction edge into square, therefore $I_{1}>I_{0}$, see Figure \ref{Fig:1}, with this in mind, 
we normalized the intensities for each one in order to observe the displacement of the spots.

We consider  $B$, $F$, and $R$ to be normalized quantities, such that $B=b/w_{0}$, $R=r/w_{0}$, and $F=f/L_{D}$, where $w_{0}$ stands for the Gaussian beam waist and $L_{D}$ is the Rayleigh distance given by $L_{D}=\pi w_{0}^{2}/\lambda$.

\begin{figure}[ht]
\centerline{\includegraphics[width=9cm]{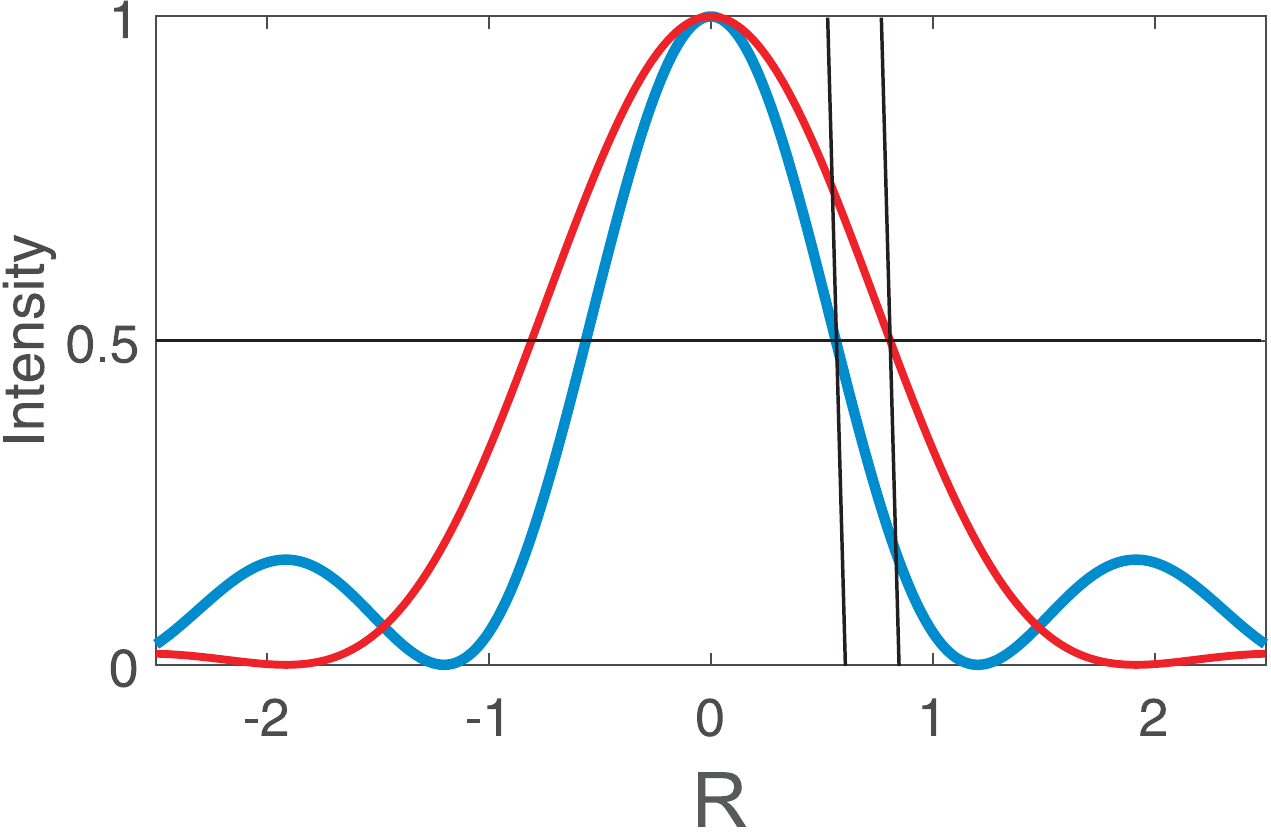}}
\caption{Transverse profile of the intensity distributions for the Arago (blue) and Airy (red) spots. Notice that the FWHM of the Airy disk is larger than in the Arago spot.}
\label{Fig:2}
\end{figure}

Figure \ref{Fig:2} shows the transverse intensity profile between the Arago and Airy distributions. The former, generated by a finite thin lens of radius $B=1$ and focal length $F=1$, and the latter, by a circular obstruction of the same radius. Notice that the Full Width Half Maximum (FWHM) of the main lobe of the Airy distribution is $0.483$ units wider than the main lobe of the Arago distribution. In the next sections, we use the main lobes for comparison. There are other ways to measure the size of spots and their displacement, e. g, the algorithms that measure the center  of spot (centroid algorithms) \cite{Leroux,Carvallo}; however, here we considering enough the of spots selection with the maxima of main lobes.


\newpage

Displacement of spots in the techniques Hartmann and Shack-Hartmann can be measured with the angle of deviation of spots for each gradient component. The gradient component in the direction $x$ and $y$ of the spot-map is given by
\begin{align}
\frac{\partial\Phi (x_i,y_j)}{\partial x} &\approx\frac{\triangle x_{i,j}}{F}, \nonumber \\
\frac{\partial\Phi (x_i,y_j)}{\partial y} &\approx \frac{\triangle y_{i,j}}{F},
\label{gradiente}
\end{align}
where $\Phi$ is the wavefront, and $\triangle x_{i,j}$ and $\triangle y_{i,j}$  are measure displacement along the  $x$ and $y$ directions of the spots of hole or lens $(i,j)$ in the position $F$ \cite{spiricon}. In the next section, we are observed the position of displacement of the spot-map with a simulation numerical of propagation of diffraction of the array of apertures and obstructions in the focal plane.

\section{Comparison between the spot-map produced by the Arago spots and the Shack-Hartmann technique}

We simulate the behavior of a plane, cylindrical, spherical, and vortex waves diffracted by an array of circular lenses and circular obstructions. We solve numerically for the paraxial wave equation  $2i\partial u/\partial z + \nabla^2 u = 0$ \cite{Arfken,jobmh}. The transverse radial coordinate is normalized to the beam waist $w_{0}$, while the propagation distance $z$ with respect to $L_{D}$. 
We consider an array of 49 lenses and circular obstructions, both with radius $r=0.25B$, and a separation among their centers of $0.5B$, the focal distance is $F=(1/150)L_{D}$, which corresponds to our image plane. The optical parameters are consistent with the data from commercial wavefront sensors, e.g. the SH sensor from Thorlabs (model WFS150-7AR or WFS10-14AR).

\begin{figure}[ht]
\centerline{\includegraphics[width=10cm]{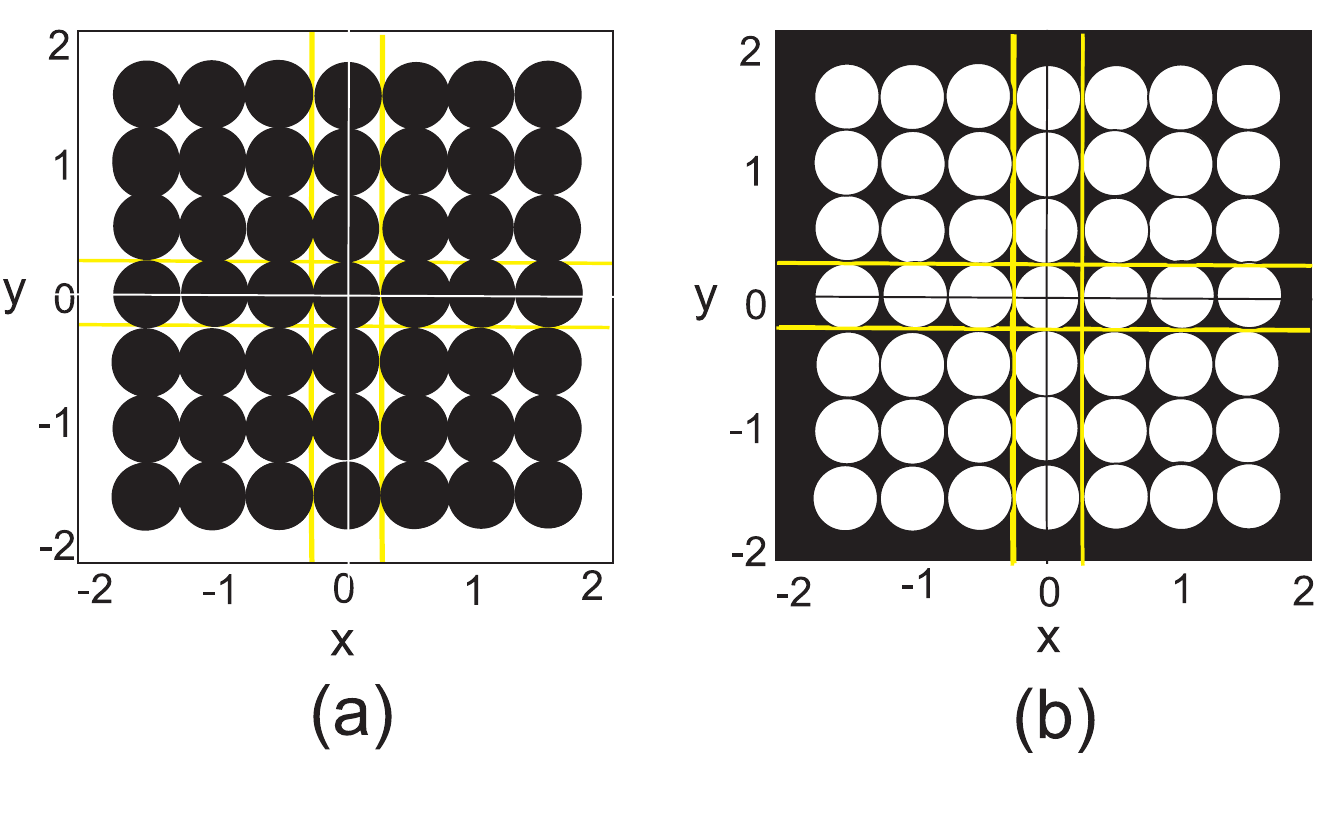}}
\caption{Schematic of the problem. (a) circular obstructions, and (b) microlenses array. The yellow lines delimit the detection area for the central element of each array.}
\label{Fig:3}
\end{figure}

Figure \ref{Fig:3} shows schematic the array of lenses and circular obstructions into computational window of $2.5 x 2.5$ B units with a number $2^{11}=2048$ pixels. The region of detection for each lens or obstruction is limited by the square that bounds each element. Figure \ref{Fig:3} shows the detection area for the central element of each array. 

\begin{figure}[ht]
\centerline{\includegraphics[width=10cm]{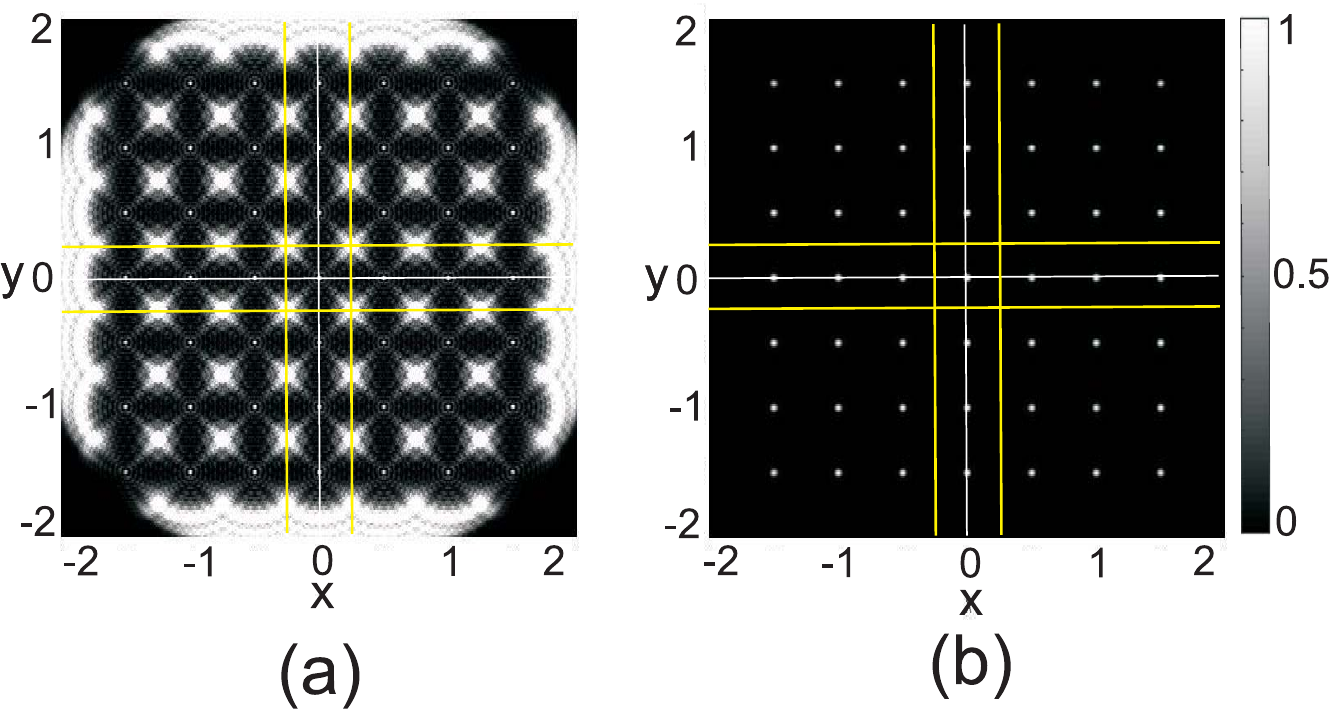}}
\caption{(a) Arago spot pattern produced by the array of circular obstructions shown in Fig. \ref{Fig:3}(a). Conversely, (b) shows the Airy spot pattern from the array of microlenses shown in Fig. \ref{Fig:3}(b).}
\label{Fig:4}
\end{figure}

The first step is to obtain the reference spot-map with a plane wave at normal incidence through the array of circular lenses and obstructions. The spot-map produced by this plane wave is shown in Fig. \ref{Fig:4}. Figure \ref{Fig:4}(a) shows the Arago spot pattern produced by the array of circular obstructions. Conversely, Fig. \ref{Fig:4}(b) shows the Airy spot pattern from the array of microlenses. 

Diffraction pattern from the obstructions shows two main features. One is the Arago spot, which is localized at the optical axis of each obstruction. The second feature is the diffracted light that passes through the apertures between the obstructions. In our numerical method, we use a circular super-Gaussian function to eliminate the reflections from the borders of our numerical window. Notice that the apodization eliminates the light at the edges of the window. However, it does not affect our region of interest.  Experimentally can be used an element circular to block the light, and the diffraction of this element arrives to the center of the distribution.  

\begin{figure}[ht]
\centerline{\includegraphics[width=8cm]{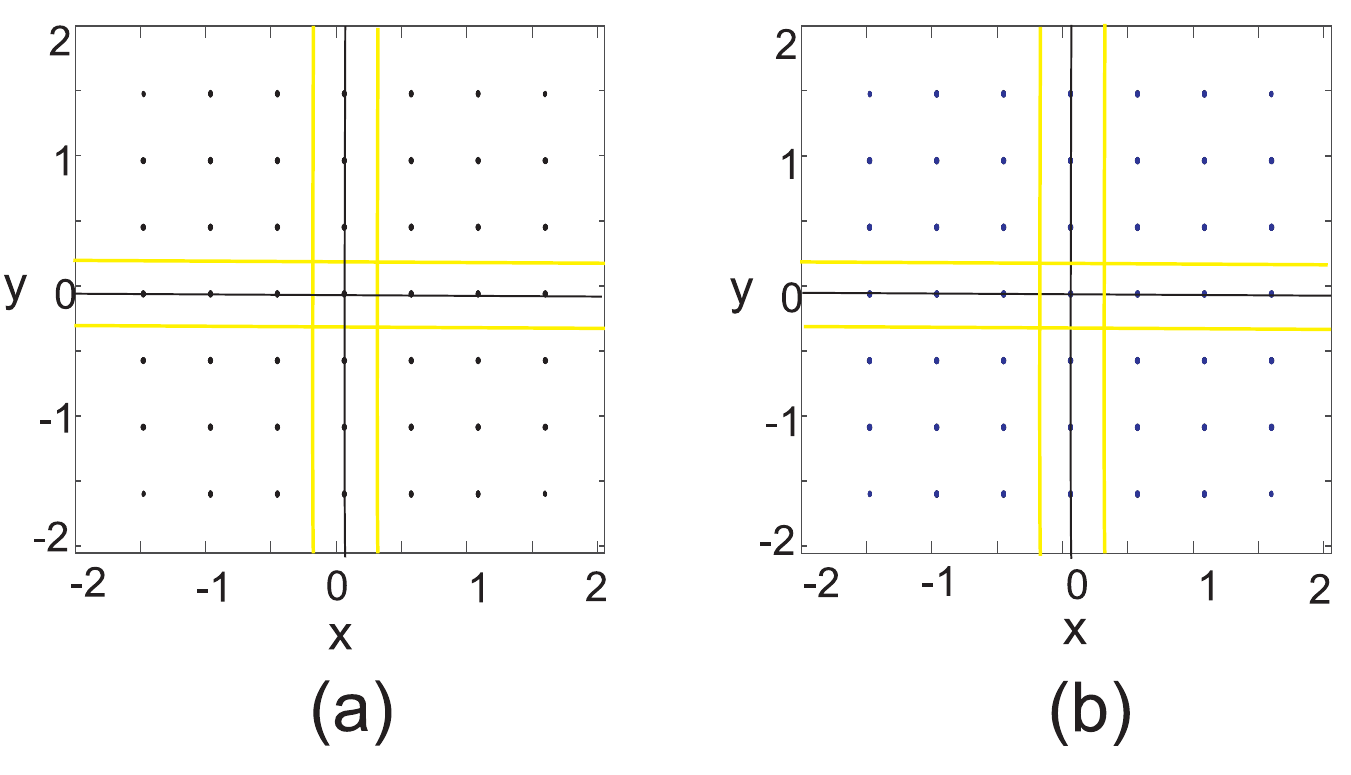}}
\caption{Spot-map in: (a) Arago technique, and (b) Shack-Hartmann technique.}
\label{Fig:5}
\end{figure}

\begin{figure}[ht]
\centerline{\includegraphics[width=9cm]{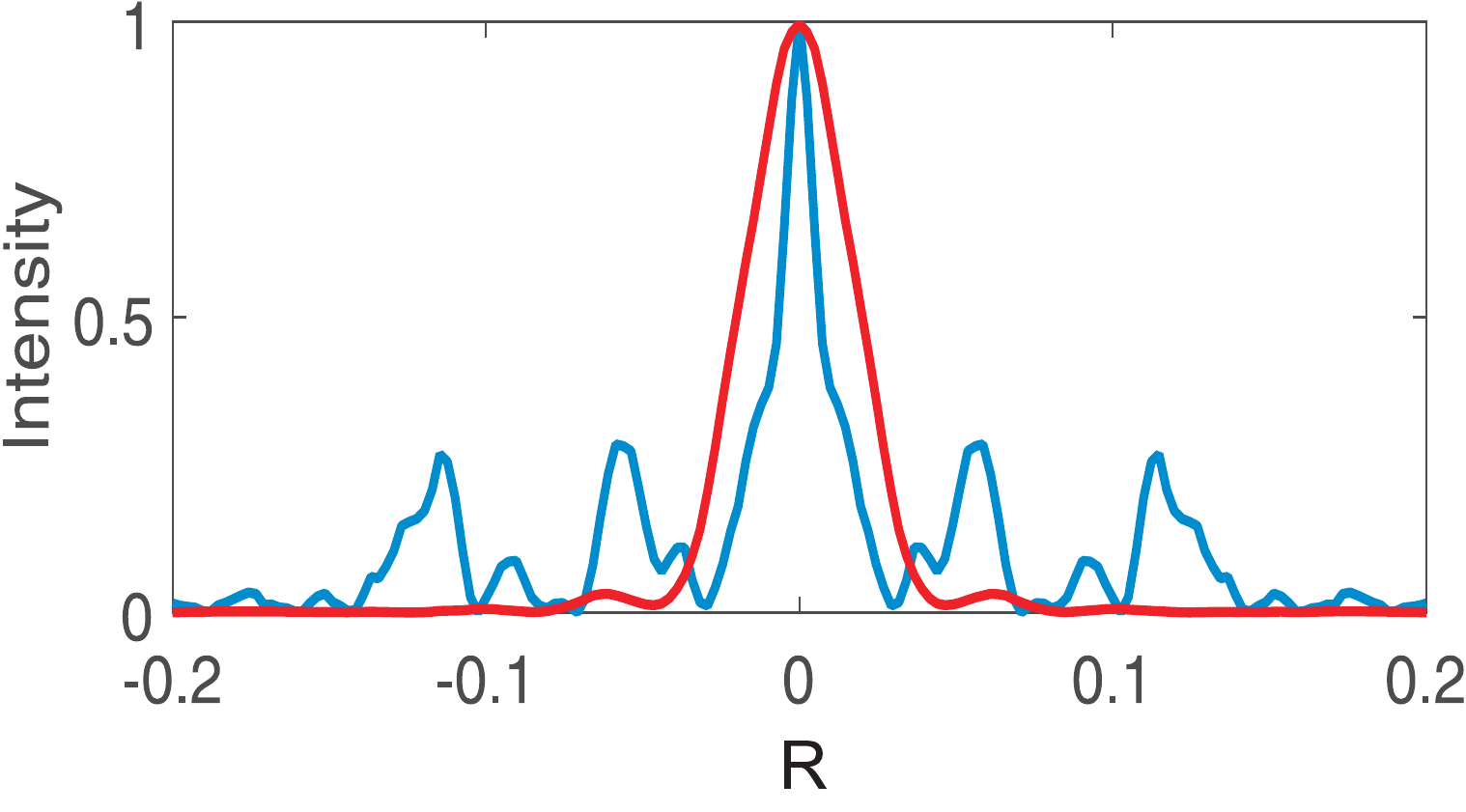}}
\caption{Transverse profile of intensity of the spot in the origin: Arago (blue), and Airy (red).}
\label{Fig:6}
\end{figure}

We selected the maxima of the Arago and Airy spots from Fig. \ref{Fig:4} to obtain the spot patterns shown in Fig. \ref{Fig:5}, and these are reference map. Figure \ref{Fig:6} shows the intensity of the transverse profile of the central spot for each distribution where the profile of the diffraction. Similar to the distribution shown in Fig. \ref{Fig:2}, the width of the Arago spot is narrower than the Airy spot. The side lobes of the Arago spot are perturbed by the diffraction of the local obstructions. Nevertheless, the central lobe is well defined and therefore, easy to localize. The spots are in the same position  therefore we considering  that both techniques can be useful for recover the wavefront. 

\begin{figure}[p]
\centerline{\includegraphics[width=10cm]{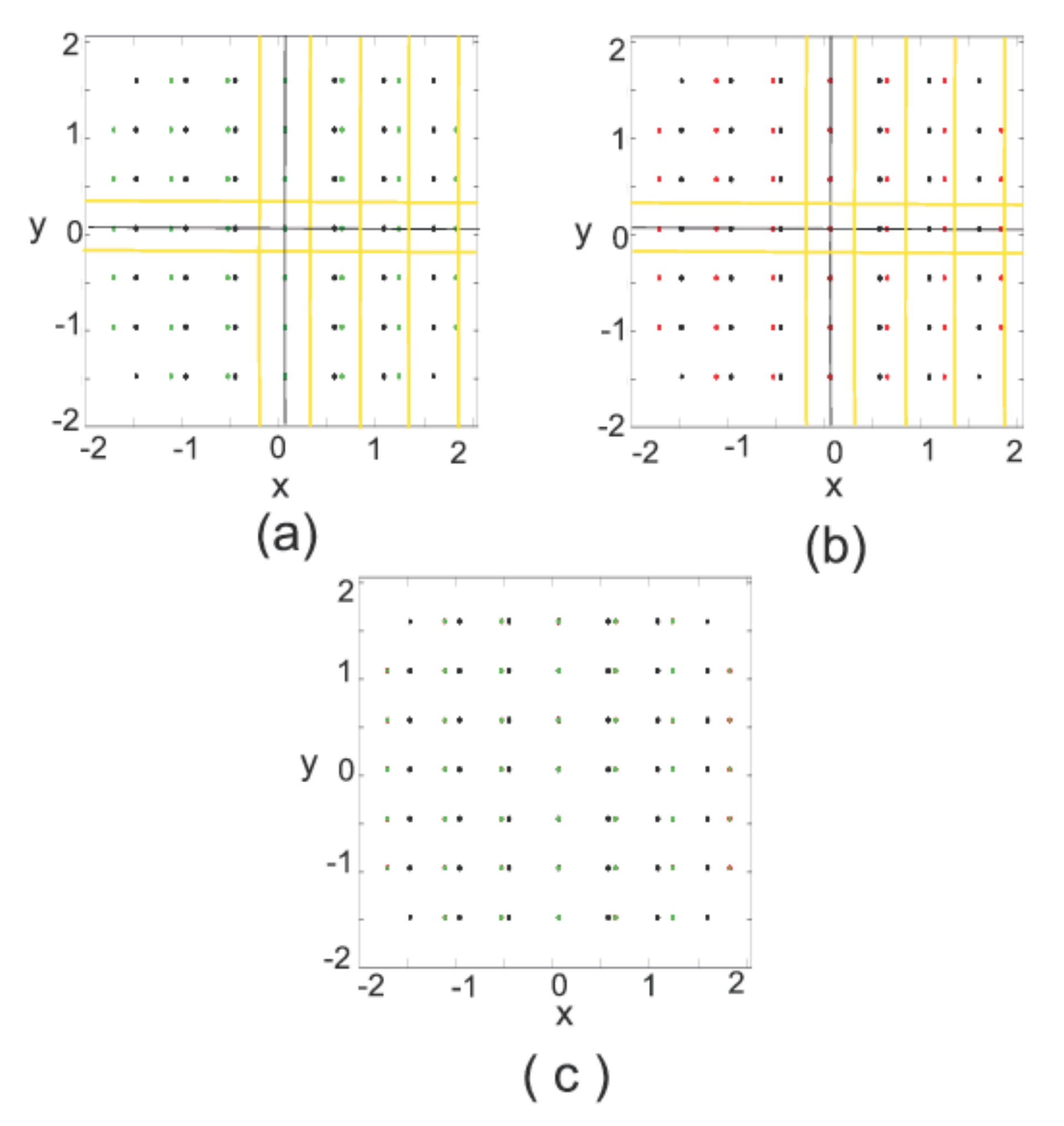}}
\caption{Spot-map for cylindrical wave. (a) Arago spot, and (b) Airy spot, and (c) both techniques. It is observed the displacement in the spots respect to the black spots of reference.}
\label{Fig:7}
\end{figure}

\newpage
After of obtaining the reference spot-map, we analyze the spot-map for different waves. The first wave that we analyzed, it is a cylindrical wave described by $U=\exp(ix^{2}/F_{c})$, with a focal length of $F_{c}=1/25$. Figure \ref{Fig:7} shows the spot-map generated by the two techniques. Now, it is possible to observe a relative displacement, in the direction $\pm x$, of the spots with respect to the reference pattern, which are represented by the black spots. Figure \ref{Fig:7}(c) shows simultaneously both distributions. It is noticed that the location of both spot maps perfectly overlap between them. The width of each spot corresponds to the FWHM and the location to the position of its maximum. 
\newpage

\begin{figure}[p]
\centerline{\includegraphics[width=11cm]{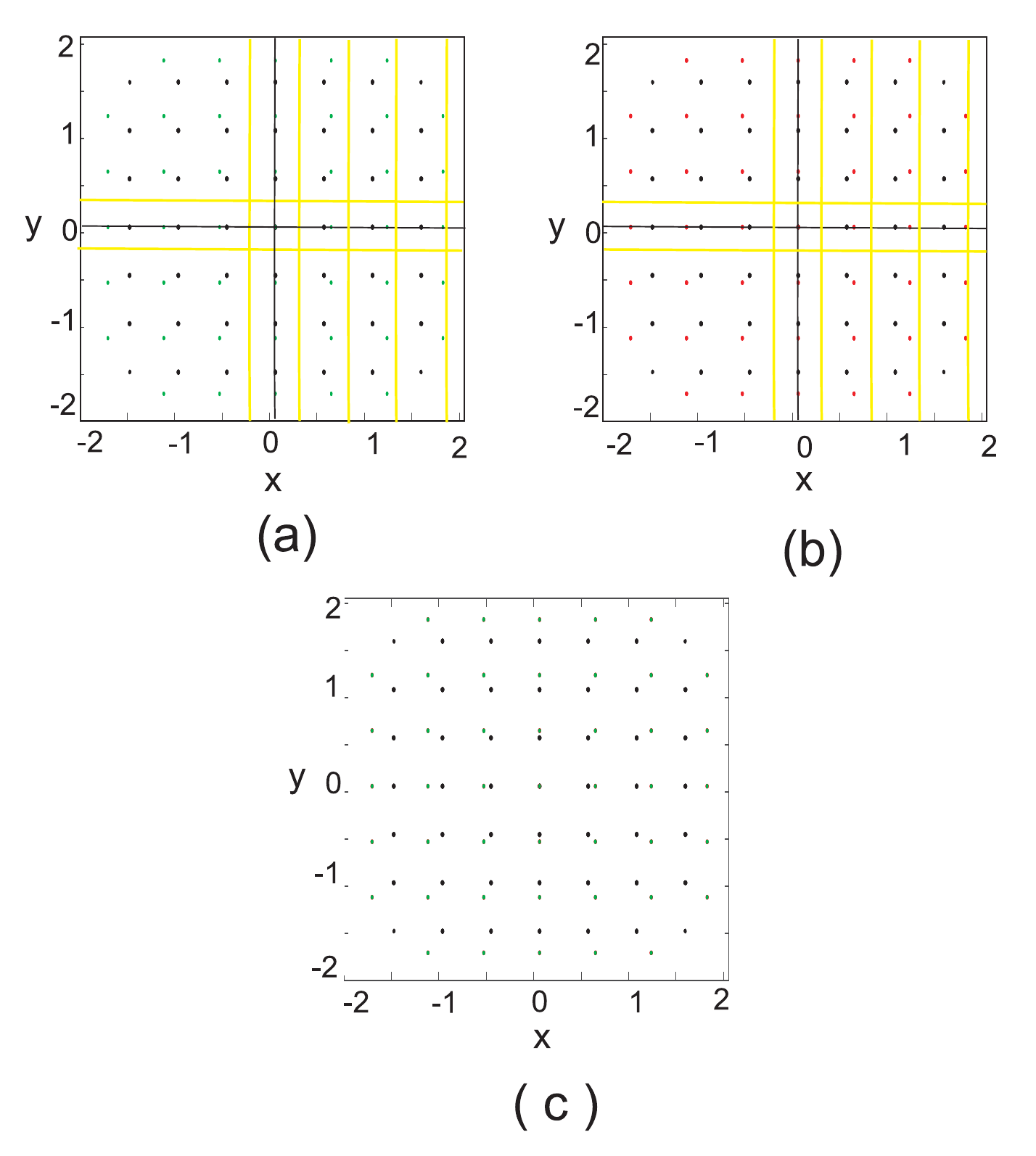}}
\caption{Spot-map for a spherical wave: (a) Arago spot, (b) Shack-Hartmann technique, and (c) both techniques. It is observed the displacement radial away of the reference spots.}
\label{Fig:8}
\end{figure}

Now, we analyzed a spherical wavefront described by $U=\exp(i(x^{2}+y^{2})/F_{sp})$, with a focal length $F_{sp}=1/25$. Figure \ref{Fig:8} shows the spot-map generated by this spherical wave. For the two techniques, the resulting spots move away from the reference black spots in a radial direction. Again, the location of both spot patterns perfectly overlap, as shown in Figure \ref{Fig:7}(c). Therefore we considering that inverted Hartmann technique can be useful for recover the wavefront in some common waves and combination of them. The next test will be with a wave non uniform and an  singularity in origin.

\newpage
Finally, we analyze a vortex wave with a topological charge $m$ described as $U=A\exp(im\phi)$. The vortex wave is characterized by a spiral wavefront and an annular intensity distribution. During the propagation, the phase rotates and the radius of the intensity changes with propagation, dictated by the paraxial propagation. 

\begin{figure}[p]
\centerline{\includegraphics[width=10cm]{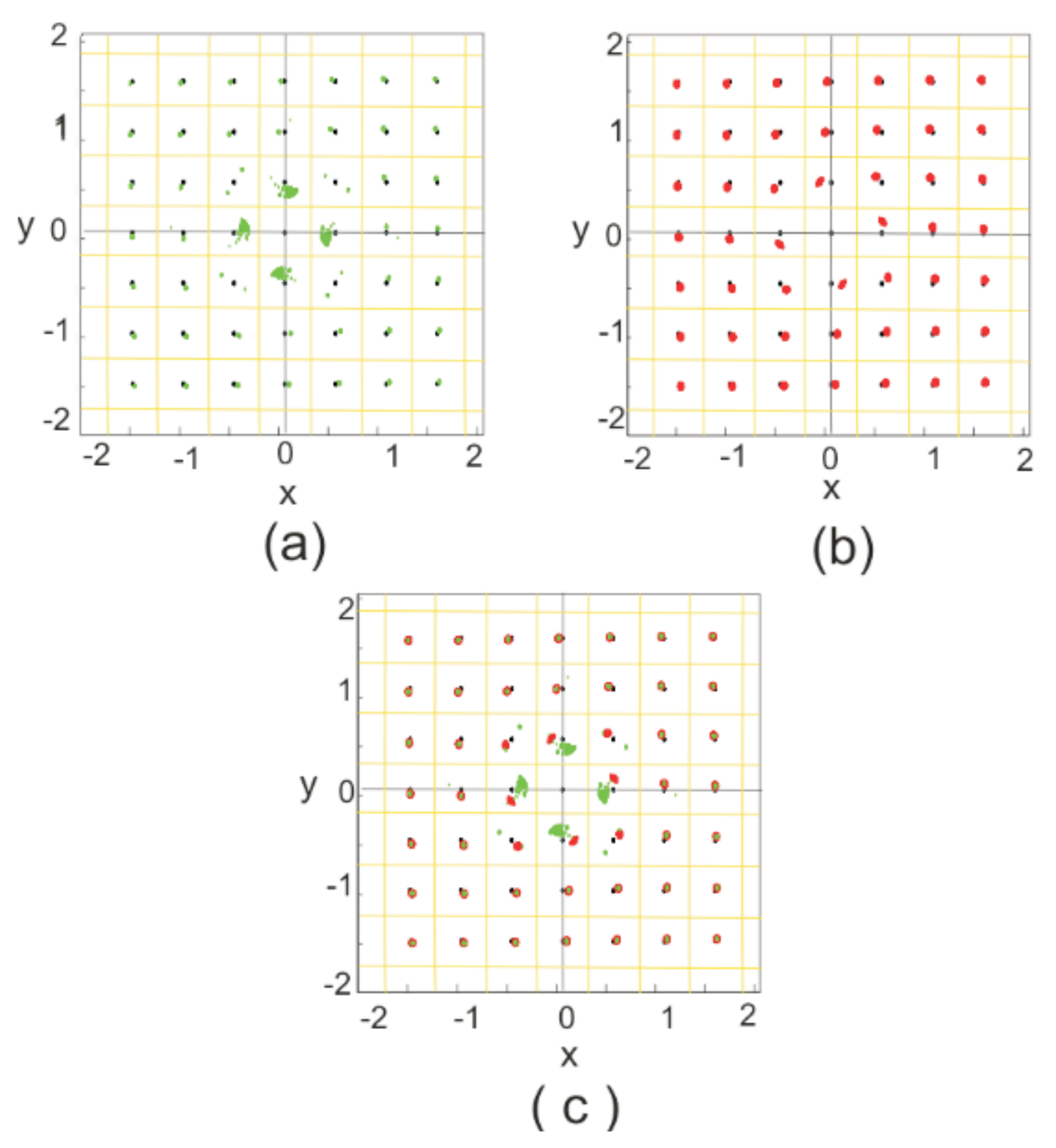}}
\caption{Spot-map for vortex wave: (a) Arago spot, (b) and Shack-Hartmann technique, and (c) both techniques. Notice that the Airy spot is larger than Arago spot.}
\label{Fig:9}
\end{figure}

In order to have a measurable change in the spot-map respect  to the reference, (size, array of lenses and obstructions, and focal length) is necessary a vortex with large charge or have large distance at propagation respect to $F$. After to do some tests with different charges $m$ and distances in the propagation, we select the charge $m=20$ in order to not modify the conditions in the arrays. The result is shown in the Figure \ref{Fig:9}.

The spot-map generated by the vortex wave has a spiral behaviour  around of the reference spot for the two techniques. To select the position of the spots, we find the points where the intensity is higher than the half maximum. However, in this case, the intensity distribution for the obstructions close to the vortex singularity shows an inhomogeneous diffraction pattern. This pattern reduces the maximum intensity of that spot below the half maximum of the other spots. Therefore, in order to detect those spots, we reduce our threshold to $0.35$ of the maximum (see Figure \ref{Fig:9}(a)). However, this selection process comes with other details. There appear additional spots within the same detection area. The Airy spot, on the contrary, does not show this spurious effects, as shown in the Figure \ref{Fig:9}(a). The Figure \ref{Fig:9}(c) shows the spot patterns for both cases.

\newpage
\section{Conclusions}
We have demonstrated that using an array of circular obstructions is possible to measure the spot-map of wavefront by observing the Arago spot pattern. The displacement from the local deviation of each spot is measured with a simulation numerical and compared with the Shack-Hartmann technique, and the both techniques show the same spot-map to recover the wavefront.

The inverted Hartmann  sensor works fine with cylindrical and spherical wavefronts, and their combinations. To helical wavefronts, the method showed spurious effects due to the diffraction from the apertures between the obstructions close to singularity. Nevertheless, the method can have technical advantages in the fabrication of circular obstructions instead of lenses,  and it can use in non-visible wavelength \cite{Bakken,Cui}, where it is used Hartmann technique and with the same results that Shack-Hartmann technique. Finally, the inverted Hartmann technique might be useful to recover the wavefront replacing the Hartmann and Shack-Hartmann techniques.

\section*{Funding}
Consejo Nacional de Ciencia y Tecnolog\'{i}a (CONACYT) (Grants: 257517, 280181, 293471).

\section*{Acknowledgments}
JMH acknowledges partial support from CONACYT, M\'{e}xico.

\end{document}